# Polarization effect and emission control in asymmetric cross-shaped slot antennas surrounded with periodic corrugations

A. Djalalian-Assl[1]

[1]51 Golf View Drive, Craigieburn, VIC 3064, Australia
E-mail address: amir.djalalian@gmail.com

## ABSTRACT

I discuss progress in the development of asymmetric cross-shaped plasmonic antennas based on resonant nanoscale apertures surrounded by surface corrugations. By tailoring the aperture and the surrounding surface, we show directionality and polarization control of transmitted light.

## 1. INTRODUCTION

The ability to control the state of polarization and the radiation pattern as well as increase the radiation rate of quantum emitters is highly desirable for next generation telecommunications and quantum computing. Nanometric apertures perforated in metallic films exhibit localized surface plasmon resonances (LSPR) which depend on the shape and size of the aperture and the dielectric environment of the film. We have previously shown computationally and experimentally that a high degree of circular polarization is achievable by detuning the two orthogonal LSP modes in an array of asymmetric cross-cavities [1,2]. The utilization of surface plasmon polaritons and their coupling to the LSP modes in an elliptical bull's eye structure was also shown to produce the same effect[3]. Apertures in metallic films surrounded by surface corrugations have also been shown to control the directionality of the emission from dipoles in apertures[4]. We discuss progress in the development of asymmetric cross-shaped plasmonic antennas based on resonant nanoscale apertures surrounded by surface corrugations. By tailoring the aperture and the surrounding surface, we show potential for the control of directionality and polarization of transmitted light.

### 1.1 Simulation

A single, isolated symmetric cross cavity aperture perforated in a 100 nm thick silver film on a glass substrate was modeled using the finite element method, implemented in COMSOL Multiphysics 4.3b. The refractive index of the glass substrate was set to $n_{glass}$=1.52, whereas the wavelength dependent refractive index for bulk silver, $n_{silver}(\lambda)$, was obtained from Palik[5]. The model was excited by a normally incident field, $\lambda$=700 nm linearly polarized at 45° to the *x*- and the *y*-axis, on the glass/silver interface and the transmitted power was calculated by integrating the time averaged Poynting vector over a constant surface area occupied by a hemisphere on the air/silver side, see Figure 1. Optimization of the cavity's arm-length, *L*, was carried out by setting the arm-width, *W*, to 40 nm while *L* was increased from 130 to 220 nm. Figure 2 depicts the transmitted power spectrum (normalized to the peak value) vs *L*. The peak at *L*=170 nm corresponds to the cavity's LSPR at $\lambda$=700 nm. For *L*=170 nm, the polar plot of the magnitude of the time averaged radial component of Poynting vector vs. the angular direction in *x-z* plane, i.e. radiation pattern [6], shows that the central lobe broadens, with increase in *z*, see Figure 3.

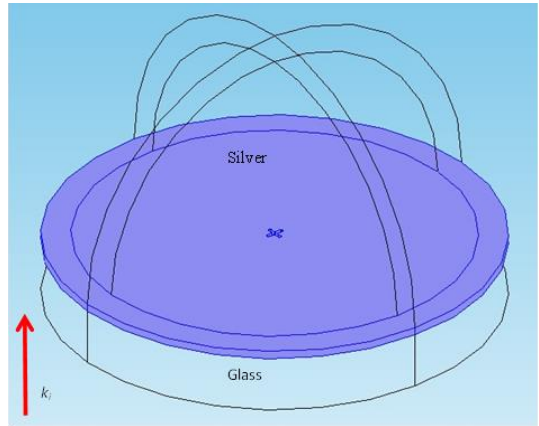

Figure 1: COMSOL Multiphysics model to examine the transmission response of a symmetric cross perforated in a 100nm thick silver film on a glass substrate.

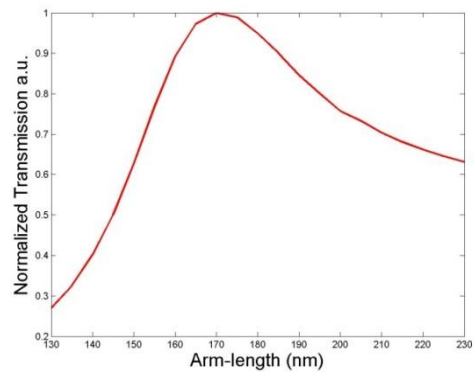

Figure 2: Calculated normalized transmission vs. the arm-length of a symmetric cross cavity aperture perforated in a 100nm thick silver film on a glass substrate. Arm-width, $W$, was fixed at 40 nm while $L$ was increased from 130 to 220 nm.

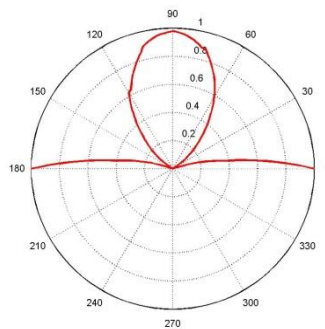

Figure 3: Calculated transmitted radiation (power) pattern of an isolated symmetric cross with arm-lengths L=170 nm and arm-width W=40 nm perforated in a 100nm thick silver film on a glass substrate.

To produce circularly polarized light (CPL) on transmission through a nano-cavity, two orthogonal modes, LSP in this case, with equal amplitude and a phase difference of $\pi/2$ are needed. Having found the resonant arm-length, $L$=170 nm at our target wavelength $\lambda$=700 nm, we can now search for the dimensions of an asymmetric cross cavity aperture in such a way that the relative phases associated to the LSPs along two arms sum up to $\pi/2$, i.e. $\Phi_x - \Phi_y = \pi/2$. Relative phases were calculated by fixing one arm-length, $L_x$=170 nm, and varying the orthogonal one $L_y$. Figure 4 shows the normalized transmission and the relative phase differences. From the result obtained, it is evident that an asymmetric cross cavity having two unequal arms, $L_1$=150 nm and $L_2$=220 nm, support two orthogonal LSP modes that oscillate with a total phase difference of $\pi/2$, when excited with an incident wave having $\lambda$=700 nm and linearly polarized at 45° to the cross arms. Note that the relative phases associated with $L_1$ and $L_2$ are $\pm\pi/2$ from the 0 phase line which is associated to the LSPR. This ensures that the optimum operating wavelength at $\lambda=700$ nm is maintained. Although 45° incident polarization guarantees the phase requirement, it does not guarantee equal oscillation strengths. Coupling between the LSPs and the incident electric field, which is also a function of the incident polarization, dictates the strength of the LSP oscillations. In other words, by varying the incident polarization, one can control the strength. To search for the optimum incident polarization, Stokes parameters vs. the incident polarization were calculated. It was found that for incident polarizations ranging from 40° to 50°, transmitted light was circularly polarized, see Figure 5.

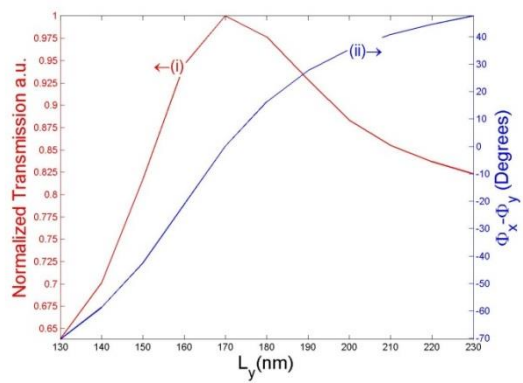

Figure 4: Transmission and phase spectra of an asymmetric cross cavity aperture perforated in a 100nm thick silver film deposited on a glass substrate vs. the change in one arm-length, $L_y$ while the other arm-length was fixed at $L_x$=170nm.

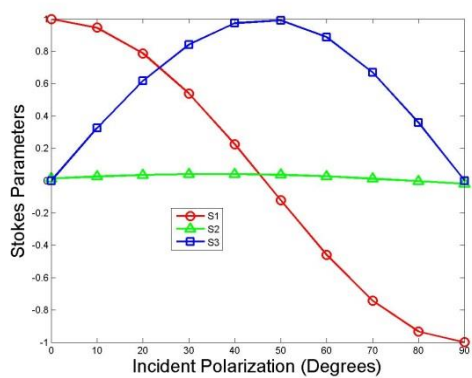

Figure 5: Stokes parameters vs. the incident polarization for an isolated asymmetric cross cavity aperture in a 100nm thick silver film on a glass substrate, with arm-lengths $L_x$=150 nm and $L_y$=220 nm.

The presence of periodic surface features gives rise to Surface Plasmon Polaritons (SPP) Bloch modes that may be excited by the LSPs in the nearby cavities [7]. Both the LSPs and the SPPs decay into free propagating EM waves. The resultant radiation pattern is due to the interaction of these phenomena. Here we propose concentric circular corrugations on air/silver interface surrounding the cavity aperture at its center. To prevent destructive interference between the two classes of surface plasmons, and hence depolarization, we first optimize the dimensions of the corrugations around a symmetric cross with arm-lengths $L$=170nm at our target wavelength, thus optimizing the corrugations around the cavity's LSPR. In doing so, we search for the optimum radius of the innermost circle, $R_{corr}$, and the optimum periodicity, $P_{corr}$, that maximizes the coupling between the cavity's LSPR and the corrugation's SPP at $\lambda=700$ nm. The width and the depth of the corrugations are fixed at $W_{corr}$=40 nm and $H_{corr}$=50 nm respectively to reduce the number of degrees of freedom in our search. Once the dimensions for the corrugations are known, we tailor the surrounding surface of the asymmetric cross accordingly and examine the transmission response. Figure 6 and Figure 7 show that the transmission through the asymmetric cross-shaped nano-aperture is at its maximum when $R_{corr}$=710 nm and $P_{corr}$=650 nm respectively. Figure 8 represents the model for the concentric circular corrugations at air/silver interface surrounding the asymmetric cross cavity aperture at its center. The radiation pattern seen in Figure 9 shows a high degree of directionality in transmitted light in comparison to that of an isolated cavity seen in Figure 3. Stokes parameters depicted in Figure 10, were calculated at the center of the cavity at $z=0$, i.e. air/silver surface. Our model shows that a high degree of CPL is produced on transmission in the 690<$\lambda$<740 nm range when excited from the glass/silver with 45° incident polarization.

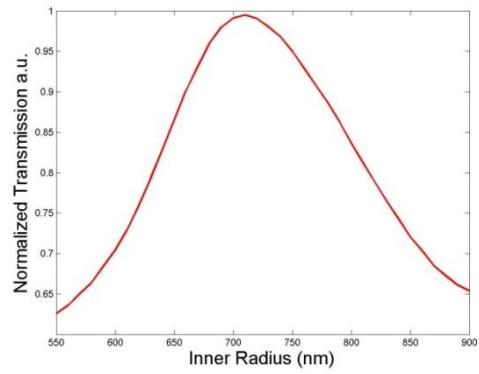

Figure 6: Transmission spectrum vs. the inner radius $R_{corr}$ of a symmetric cross cavity aperture perforated in a 100nm thick silver film on a glass substrate surrounded by periodic corrugations.

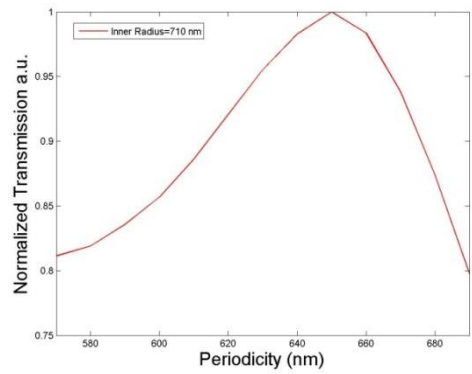

Figure 7: Transmission spectrum vs. the periodicity $P_{corr}$ of a symmetric cross cavity aperture perforated in a 100nm thick silver film on a glass substrate surrounded by periodic corrugations.

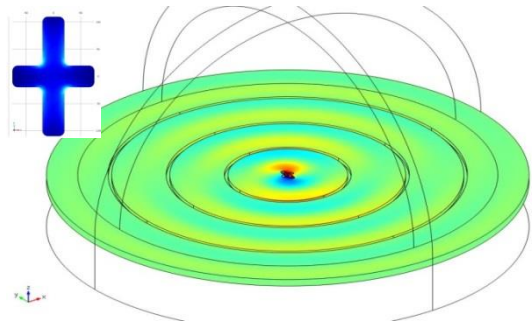

Figure 8: Corrugations implemented in the surrounding surface of the asymmetric cross cavity aperture in silver film. Corrugation dimensions: $W_{corr}$=40 nm, $H_{corr}$=50 nm $R_{corr}$=710 nm and $P_{corr}$=650 nm.

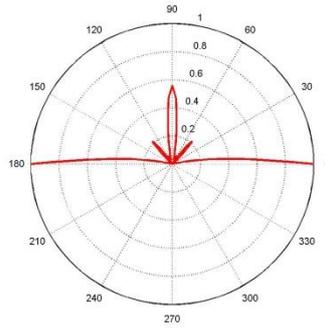

Figure 9: Radiation (power) pattern of a single cross cavity aperture with arm-lengths $L_x$=150 nm and $L_y$=220 nm surrounded by concentric circular corrugations with dimensions, $W_{corr}$=40 nm, $H_{corr}$=50 nm $R_{corr}$=710 nm and $P_{corr}$=650 nm.

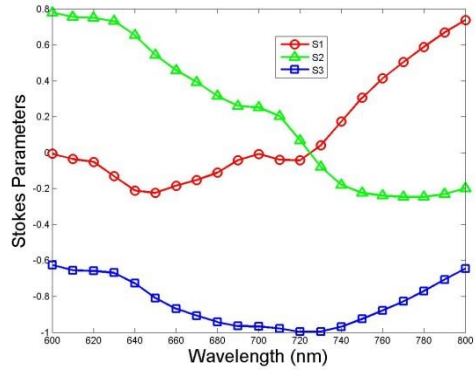

Figure 10: Stokes parameters calculated at the center of the asymmetric cross at $z$=0, i.e. air/silver surface. High degree of transmitted light with high degree of circular polarization is produced in the 690<$\lambda$<740 nm range when excited from the glass/silver with 45° incident polarization.

## 1.2 Fabrication and Characterization

Film preparation was carried out by first depositing a 2 nm thick germanium film (as an adhesion layer) on a glass substrate followed by a 100 nm thick silver film using IntlVac Nanochrome II electron beam evaporator. To verify the validity of the numerical results, a series of isolated symmetric cross cavity apertures with various arm-lengths were milled using a Helios NanoLab 600 Focused Ion Beam (FIB). Crosses were characterized using an unpolarized Nikon IT-U microscope halogen light source in dark-field mode. The experimental results, as seen in Figure 11, shows that at our target wavelength, $\lambda$=700 nm, the LSPR occurs for a symmetric cross cavity aperture with arm-lengths $L$≈210 nm. The discrepancy in the arm-lengths obtained numerically versus those obtained experimentally, may be attributed to the use of bulk silver refractive indices which no longer hold in nanostructure materials. Other factors such as the film's granularity and the anisotropic profile of the cavity may also contribute to this inconsistency. As a consequence a +40 nm increase in arm-length must be taken into account when fabricating cavities based on the dimensions obtained numerically. A single asymmetric cross with an arm-width $W$=55 nm and arm-lengths $L_x$=200 and $L_y$=270 nm was milled in a 100 nm thick silver film deposited on glass substrate, see Figure 12. The artefacts associated with sidewall nonuniformity due to the milling process, also contributes to the fabrication error. For example the width of the cross measures $W$=55 nm at $z$=0, i.e. the air/silver surface, but reduces to $W$=35 nm at $z$= -100 nm, i.e. the cavity walls are sloped. Concentric corrugations with $R_{corr}$=725 nm and $P_{corr}$=620 nm were milled in the surrounding surface, see Figure 13. Note that the fabricated dimensions differ from the intended dimensions, i.e. $R_{corr}$ =710 nm and $P_{corr}$=650 nm, due to the intrinsic errors associated with the FIB milling process. The position of the cross within the concentric corrugation is also slightly off-center, see Figure 14. Figure 15 shows the image of the device, captured using an optical microscope in transmission mode. The presence of a bright spot at the center and the luminescent rings surrounding it attest to the excitation of LSPR, and consequently, SPPs despite the fact that the device was illuminated from the glass/silver side.

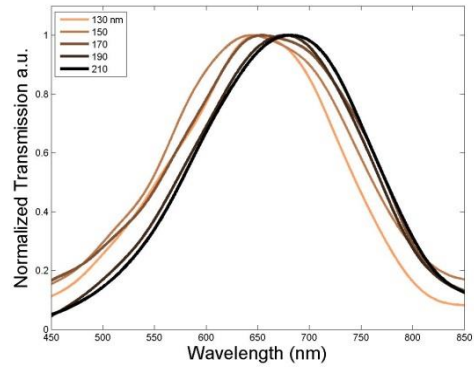

Figure 11: Transmission spectra for a set of isolated symmetric crosses with various arm-lengths perforated in a 100 nm thick silver film deposited on a glass substrate.

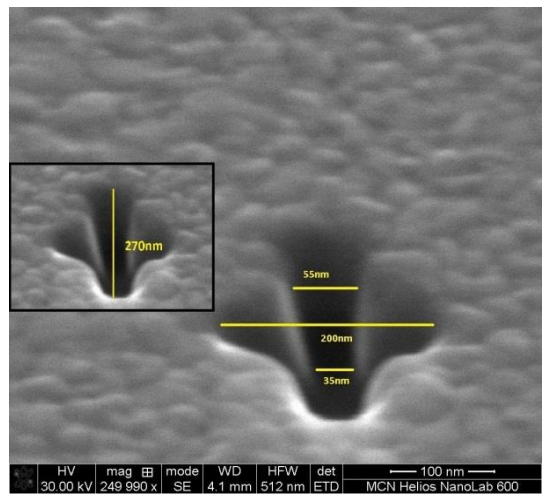

Figure 12: SEM image of the asymmetric cross cavity aperture. $L_x$=200 nm $L_y$=270 nm (inset). Anisotropy in the vertical direction is noticeable, i.e. arm-width, $W$, changes from 55 nm near the air/silver interface to 35 nm near the glass/silver interface.

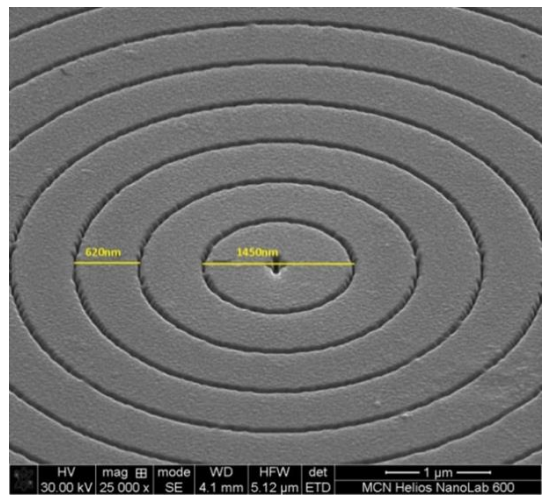

Figure 13: Concentric corrugations surrounding the asymmetric cross. The radius of the inner circle $R_{corr}$=725 nm and the periodicity $P_{corr}$=620 nm.

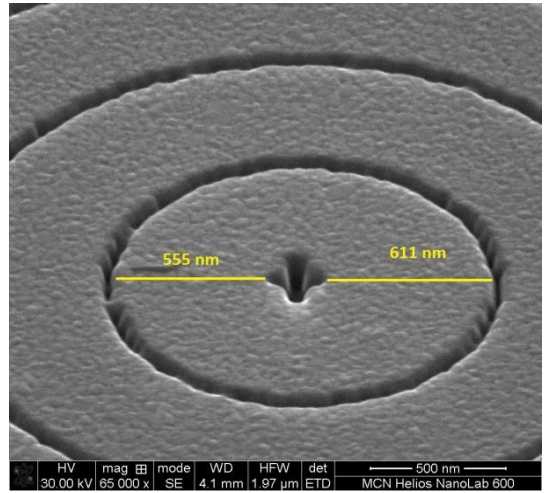

Figure 14: position of the cross is slightly off-centered with respect to the concentric corrugations.

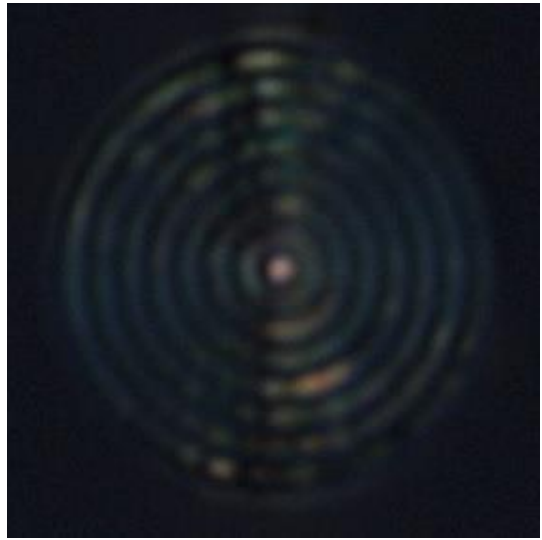

Figure 15: Captured image using an optical microscope in transmission mode. Presence of a bright spot at the center and luminescent rings surrounding it attest to the excitation of LSPR, and consequently, SPPs despite the fact that the device was illuminated from the glass/silver side.

The first step in characterizing the device was to measure the background florescence that may result from gallium ions being implanted into the cavity or corrugation sites during the milling process. Using a confocal set up, the air/silver interface of the device was illuminated with a 500μW green laser, λ=532nm, focused through a LU PLAN Nikon 100x 0.95 objective. The reflected light from the device was detected by the same objective and passed through a high-pass λ>550 nm and a single band pass 650-750 nm OD6 filters and collected by a pinhole aperture coupled to a fiber splitter. The two outputs of the fiber splitter were connected to a single photon detector and a spectrometer respectively. The output of the photon detector was fed to the PC running LabView which, besides recording the data, also controlled the 3D piezoelectric translation stage on which the sample was mounted on. Note that we assume the *x* and the *y* axes to be aligned with $L_x$ and $L_y$ respectively and the optical axis of the objective, being orthogonal to the *x-y* plane, defines the *z* axis. Also the pixel size, which is defined by the displacement performed by the piezoelectric stage in the *x* and the *y* directions, was set to 100 nm. A pixel by pixel scan in the *x-y* plane around the area occupied by the device showed high background florescence, represented by bright regions that confined to certain parts of the corrugations, see Figure 16. The background florescence from the cavity site is moderately low; therefore the device is ideal for incorporating a single photon emitter inside the cavity.

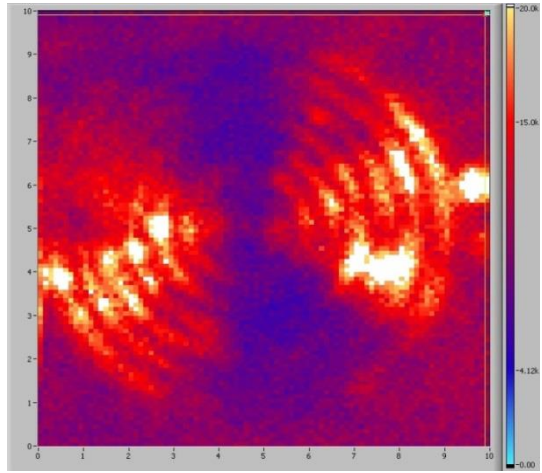

Figure 16: Background count in the 650-750 nm range, when the device was excited with a 500μW green laser, λ=532nm. White color=20k, cyan=0.

To measure the transmission spectra, the device was illuminated from the substrate side with a collimated incandescent (halogen) light, passed through a ThorLabs LP-VIS linear polarizer before incidence on the glass/silver interface. The output of the spectrometer was fed to a PC running Princeton WinSpec software. Figure 17 depicts the spectra measured when the incident polarization was set to 0°, 45° and 90°. The peak at λ=715 nm which is present in all cases, is attributed to both the SPP Bloch mode associated to the periodic corrugations and the LSPR associated with the shorter arm of the asymmetric cross. In an ideal scenario the center wavelength of the two LSP modes should coincide with the SPP Bloch mode to prevent depolarization. The overlap of the SPP mode with just one of the LSPR was unintentional and resulted from fabrication inaccuracy. The peak at λ=800 nm is attributed to the LSPR associated with the longer arm. The redshifts in center wavelength of the two LSP modes from our target 700 nm to 750 nm, and to a lesser extend in SPP Bloch mode from 700 nm to 715 nm, presents a less than ideal situation. To determine the Stokes parameters of the device vs. the wavelength, a series of intensity measurements were carried out with incident polarization set to 0°, 45°, 90° 135°, RCP and LCP. For each incident polarization, the transmitted intensity was then measured at 0°, 45°, 90° 135°, RCP and LCP. A total of 36 measurements resulting from input/output polarization permutations were obtained. Figure 18 shows the S3 parameter reaching 0.28 at around λ=715 nm when the device was illuminated with a 45° linearly polarized light.

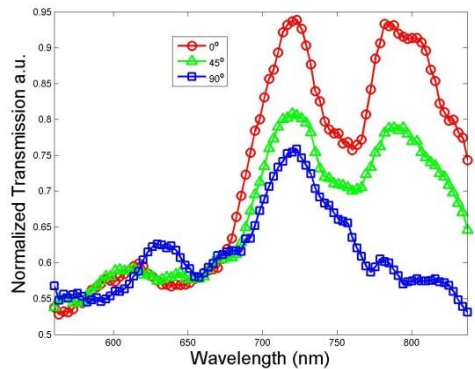

Figure 17: Transmission spectra measured for incident polarizations 0°, 45° and 90°.

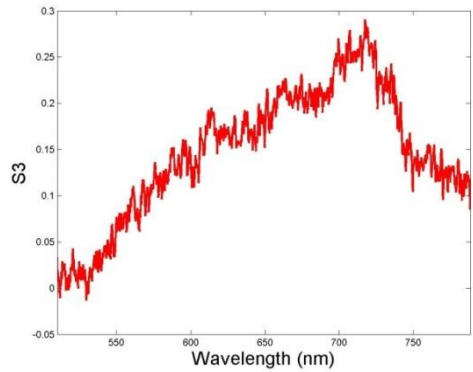

Figure 18: S3 parameter when the device was illuminated with a 45° linearly polarized light.

To examine the transmitted beam profile through the device, a series of images in the *x-y* plane were scanned by varying the distance between the sample and the objective, i.e. along the *+z* axis. The device was illuminated from the substrate side by a halogen light source filtered at λ=700 nm using a Thorlabs FB700-10 filter and transmission intensity measured at the air/silver interface using the confocal objective described above with no filter. LabView's built in algorithm for maximizing the photon count set the initial *z* position. The first image was scanned at this position and for the consecutive images the *z* was increased in 0.1 μm steps. The 2D images obtained in *x-y* plane were stacked in order of increasing *z* to form a 3D image. The resulting 3D image was then sliced in the *z-x* and *z-y* planes to obtain the transmitted beam profile. The process was repeated using at λ=750 nm using a Thorlabs FB750-10 filter. See Figure 19-22. The beam profile shows high directionality of the transmitted light.

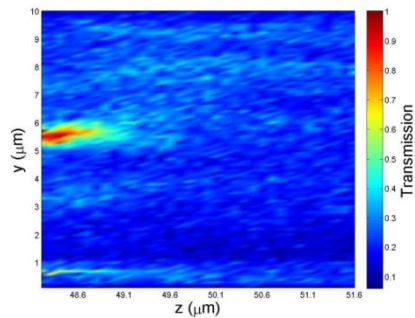

Figure 19: *y-z* slice of transmitted intensity when the wavelength of the incident light is λ=700nm.

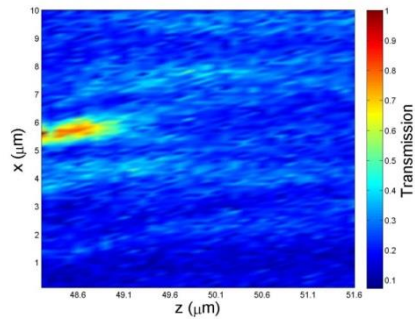

Figure 20: *x-z* slice of transmitted intensity when the wavelength of the incident light is λ=700nm.

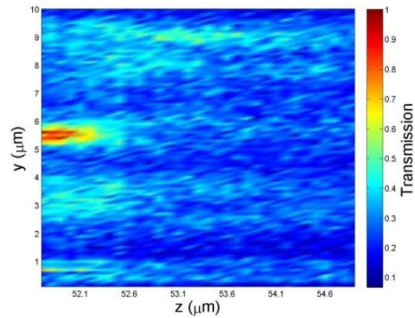

Figure 21: *y-z* slice of transmitted intensity when the wavelength of the incident light is λ=750 nm.

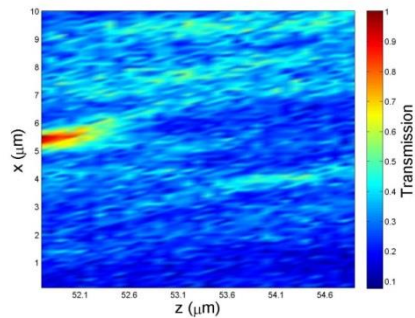

Figure 22: *x-z* slice of transmitted intensity when the wavelength of the incident light is λ=750 nm.

## 1.3 Conclusion

The transmission response of an asymmetric cross cavity aperture was numerically modeled in isolation and in conjunction to periodic surface corrugations surrounding it. Numerical results showed a high degree of circularly polarized transmitted light is achievable and the presence of concentric periodic corrugations surrounding the cross influences the directionality of the transmitted beam. An attempt in fabricating the device to the design specification presented challenges which resulted in divergence of the fabricated dimensions from those intended. Despite the differences, our experimental results confirm that it is possible to control both the directionality and polarization of the transmitted light through a metallic aperture by tailoring its dimensions and surrounding surface. Improving the device performance is the subject of ongoing work.

## 1.4 Acknowledgement